\documentclass[12pt]{article}
\usepackage{graphicx}
\usepackage{amsmath}
\def\br(#1,#2){\left\langle#1#2\right\rangle}
\def\sq(#1,#2){\left[#1#2\right]}
\def\s(#1,#2){s_{#1 #2}}
\def\t(#1,#2,#3){s_{#1 #2 #3}}

\begin{document}

\begin{titlepage}

\hspace*{\fill}\parbox[t]{5cm}
{
\today
\\MPP-2012-4} \vskip2cm
\begin{center}
{\Large \bf Constraints on Non-standard Top Quark Couplings} \\
\medskip
\bigskip\bigskip\bigskip\bigskip
{\bf Cen Zhang,$^1$ Nicolas Greiner,$^{1,2}$ and Scott Willenbrock$\,^1$}\\
\bigskip\bigskip\bigskip
$^1$ Department of Physics, University of Illinois at Urbana-Champaign \\ 1110 West Green Street, Urbana, IL  61801 \\
\bigskip
$^2$ Max-Planck-Institut f\"ur Physik,
F\"ohringer Ring 6, 80805 M\"unchen, Germany

\end{center}

\bigskip\bigskip\bigskip

\begin{abstract}
We study non-standard top quark couplings in the effective field theory approach.
All nine dimension-six operators that generate anomalous couplings
between the electroweak gauge bosons and the third-generation quarks are included.
We calculate their contributions at tree level and one loop
to all major precision electroweak observables.
The calculations are compared with data to obtain constraints on eight of these operators.
\end{abstract}

\end{titlepage}

\section{Introduction}\label{sec:intro}

Top quark interactions could provide relevant information on physics beyond
the Standard Model (SM).
Anomalous top quark interactions at colliders have been studied in the literature (see, {\itshape e.g.}
\cite{Aguilar,Tim,Bohdan,Grzadkowski:2003tf,linear1,linear2,4fermi,Cao:2007ea,Cao:2006pu,Whisnant:1997qu,Pomarol:2008bh,Scott,Degrande:2010kt}
and references therein).
In particular, a model-independent approach based on a low-energy
effective field theory
is used to describe possible new physics effects.
In this approach, high-scale physics is integrated out to obtain effective interactions
that involve only the SM particles. These interactions are suppressed by inverse powers of $\Lambda$, the scale at which the new physics resides.

Such a field theory must satisfy the $SU(3)_C\times SU(2)_L\times U(1)_Y$ symmetry of the SM.
 With this requirement, the only possible dimension-five operator violates lepton number
conservation and is irrelevant to top quark physics \cite{Weinberg:1979sa}. Thus the leading effects are generated by operators of dimension-six:
\begin{equation}
{\mathcal{L}_{{\rm eff}}=\mathcal{L}_{{\rm SM}}}+\frac{1}{\Lambda^2}\sum_i(C_iO_i+\mathrm{h.c.})
,
\end{equation}
where $O_i$ are the dimension-six operators and $C_i$ are dimensionless coefficients.
A complete list of dimension-six operators was given in \cite{Buchmuller:1985jz,Leung:1984ni,Arzt:1994gp}.
Subsequently it was found that several of this operators are not independent \cite{Grzadkowski:2003tf,AguilarSaavedra:2008zc}.
A list of 59 independent dimension-six operators is given in \cite{Grzadkowski:2010es}.

New physics that affects top-quark interactions can be parametrized using these operators.
In particular, consider anomalous top quark couplings, such as the
$W\bar{t}b$, $Z\bar{t}t$ and $\gamma\bar{t}t$ vertices. There are nine dimension-six operators that can modify
the couplings of the third-generation quarks to the $W$, $Z$ and $\gamma$ bosons:
\begin{eqnarray}
O_{\phi q}^{(3)}&=&i(\phi^\dagger\tau^I D_\mu\phi)(\bar{q}\gamma^\mu\tau^Iq)\label{op1},\\
O_{\phi q}^{(1)}&=&i(\phi^\dagger D_\mu\phi)(\bar{q}\gamma^\mu q),\\
O_{\phi t}&=&i(\phi^\dagger D_\mu\phi)(\bar{t}\gamma^\mu t),\\
O_{\phi b}&=&i(\phi^\dagger D_\mu\phi)(\bar{b}\gamma^\mu b),\\
O_{\phi\phi}&=&i(\tilde{\phi}^\dagger D_\mu\phi)(\bar{t}\gamma^\mu b),\\
O_{tW}&=&(\bar{q}\sigma^{\mu\nu}\tau^It)\tilde{\phi}W_{\mu\nu}^I,\\
O_{bW}&=&(\bar{q}\sigma^{\mu\nu}\tau^Ib)\phi W_{\mu\nu}^I,\\
O_{tB}&=&(\bar{q}\sigma^{\mu\nu}t)\tilde{\phi} B_{\mu\nu},\\
O_{bB}&=&(\bar{q}\sigma^{\mu\nu}b)\phi B_{\mu\nu},\label{op9}
\end{eqnarray}
where $q$ is the third-generation left-handed quark doublet, $t$ and $b$ are the right-handed top and bottom,
$\phi$ is the Higgs boson doublet,
$\epsilon=\left(\begin{array}{cc}0&1\\-1&0\end{array}\right)$,
$\tilde{\phi}=\epsilon\phi^*$, and
$D_\mu=\partial_\mu-i\frac{g}{2}\tau^IW_\mu^I-ig'YB_\mu$
is the covariant derivative where $\tau^I$ denote the Pauli matrices.
$W_{\mu\nu}^I=\partial_\mu W_\nu^I-\partial_\nu W^I_\mu+g\epsilon_{IJK}W^J_\mu W^K_\nu$ and
$B_{\mu\nu}=\partial_\mu B_\nu^I-\partial_\nu B^I_\mu$ are the field strength tensors of the $W$ and $B$ field.
The contribution of these operators to the vertices can be found in \cite{AguilarSaavedra:2008zc}.

Naively, from dimensional analysis we may expect that the effects of these operators are suppressed by $E^2/\Lambda^2$
where $E$ is the energy scale of the process.
This is not the case for the operators listed above in Eqs.~(\ref{op1})-(\ref{op9}).
The anomalous couplings generated by these operators violate the $SU(2)_L$ symmetry,
so they are related to the Higgs vacuum expectation value $v$.
Instead of $E^2/\Lambda^2$, these anomalous vertices scale as $v^2/\Lambda^2$,
which is independent of the energy scale of the process.
This can be seen in \cite{AguilarSaavedra:2008zc}, where the relation between
the anomalous couplings and the dimension-six operators are given.

The consequence of this scaling is that the effects of new physics will not increase with
energy. On the other hand, the effects
will not disappear in the low energy limit.
Therefore an important question is whether it is possible to extract better bounds from
electroweak precision measurements for these operators, than from the
measurements performed at high-energy colliders.

The electroweak precision measurements have a much cleaner background than hadron colliders,
and therefore are performed with a higher level of precision.
However, most of the operators listed above do not directly contribute to these measurements at tree-level.
Their corrections to the $W$, $Z$ and $\gamma$ self-energies occur at loop-level, so they
are suppressed by $g^2/(4\pi)^2$. Still, the large mass of the top quark can lead to an enhancement of the
loop-level contribution, and as a result
the constraints on top-quark anomalous couplings obtained from precision measurements may be comparable with those
obtained from collider experiments.
The top quark plays an important role as a virtual particle in precision electroweak physics.
Indeed, the correct range for the top-quark mass was anticipated by precision electroweak studies.
Now that the top-quark mass is accurately known from direct measurements, we can ask what the precision electroweak measurements have to say about the presence of dimension-six operators in loop diagrams involving the top quark.

The easiest way to put constraints on these operators is to consider the ``oblique parameters'' \cite{Peskin:1990zt,Peskin:1991sw,Barbieri:2004qk},
as most of the operators contribute only through corrections to the self-energies of the
electroweak gauge bosons. However, as we will see in Section 2, this approach is not appropriate
for all nine operators. Therefore we explicitly calculate the effects of these operators on all
electroweak measurements. We compare the results with data and perform a global fit to obtain
one-sigma bounds on the coefficients of these operators. In order to be specific,
we assume that these operators (plus two additional operators) are the only new physics effects in the theory.

The operators in Eqs.~(\ref{op1})-(\ref{op9}) may be expressed in any basis that is convenient. We choose a basis in which all fermion fields are mass eigenstates, with $q=(t_L,V_{tb}b_L+V_{ts}s_L+V_{td}d_L)$ \cite{Grzadkowski:2008mf}.  In $W$ boson self-energy diagrams, one must sum over all charge -1/3 quarks.  These quarks are much lighter than the top quark in the loop, so it is a good approximation to neglect their masses, in which case the CKM factors add up to unity.  The only place that a CKM factor survives is in photon and $Z$ boson self energies involving a $b$-quark loop and the operators $O_{bW}$ or $O_{bB}$.  These diagrams yield a factor of $V_{tb}$, which is very close to unity for three generations, and can be ignored.  The operators in Eqs.~(\ref{op1})-(\ref{op9}) also give rise to nonstandard effects such as flavor-changing neutral currents, right-handed charged currents, {\it etc.}
\cite{delAguila:2000aa}.  We do not explore the constraints on the operators from bounds on these processes.

The paper is organized as follows.
In Section 2, we discuss the oblique parameters and use them to constrain one operator as an illustration.
In Section 3, we show all major precision electroweak measurements that we will use to obtain bounds.
In section 4, we calculate the corrections to all observables from these operators, and perform the global fit.
We present our conclusions in Section 5.
Finally, we show the self-energy corrections from each operator in Appendix \ref{app:a},
and some numerical results of the global fit in Appendix \ref{app:b}.

\section{Oblique parameters}
The operators listed in Eqs.~(\ref{op1})-(\ref{op9}) affect the precision electroweak measurements in two different ways:
\begin{itemize}
\item {$O_{\phi q}^{(3)}$, $O_{\phi q}^{(1)}$ and $O_{\phi b}$
 modify the $Z\rightarrow b\bar{b}$ measurements at LEP at tree-level.}
\item {All operators modify the self-energies of $W$, $Z$ and $\gamma$ at loop-level,
and therefore affect all measurements indirectly.}
\end{itemize}
The first effect is equivalent to a correction to the $Zb\bar{b}$ couplings:
\begin{eqnarray}
&&\delta g_L^b=-\frac{1}{2}\frac{v^2}{\Lambda^2}\left(C^{(3)}_{\phi q}+C^{(1)}_{\phi q}\right),\label{eq:Zbb1}\\
&&\delta g_R^b=-\frac{1}{2}\frac{v^2}{\Lambda^2}C_{\phi b},\label{eq:Zbb2}
\end{eqnarray}
where $g_L^b$, $g_R^b$ are the left- and right-handed $Zb\bar{b}$ couplings.
This correction is easily included in the calculation.

\begin{figure}[hbt]
\centering
\includegraphics[scale=1]{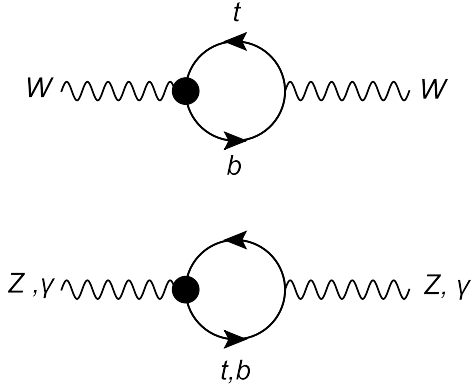}
\caption{Corrections to gauge boson self-energy. The black dots indicate the dimension-six vertex.\label{fig1}}
\end{figure}

Operators listed in Eqs.~(\ref{op1})-(\ref{op9}) modify the gauge boson self-energies through the loop diagrams shown in Figure \ref{fig1}.\footnote{There is also a diagram contributing to the $W$-boson self energy, with a top-quark loop, constructed from the
four-point contact interaction given by $O_{tW}$ and $O_{bW}$.  Since this interaction is antisymmetric in $\mu,\nu$, this diagram does not contribute to the self energy.}
They affect all quarks and leptons universally through gauge boson self-energies.
This kind of effect is referred to as ``oblique''.
Traditionally three parameters, $S$, $T$ and $U$, are used to describe the oblique new physics \cite{Peskin:1990zt,Peskin:1991sw}.
The idea is to Taylor-expand the four self-energies $\Pi_{WW}$, $\Pi_{ZZ}$, $\Pi_{\gamma\gamma}$ and $\Pi_{\gamma Z}$,
which only include the new physics contributions,
to order $q^2$.
Requiring the photon to be massless, $\Pi_{\gamma\gamma}$ and $\Pi_{\gamma Z}$ must be zero at $q^2=0$, so there will be six non-zero coefficients. Three of them are absorbed in the definition of $g$, $g'$ and $v$.
This leaves three independent parameters. The definitions of the $S$, $T$ and $U$ parameters
are given in \cite{PDG}.

We list the contribution of the nine dimension-six operators to the gauge boson self-energies in Appendix \ref{app:a}.
The coefficients $C_i$ may be taken to be
real because an imaginary part of $C_i$ violates $CP$ and will
not contribute to any self-energy.
It is straightforward to calculate the three oblique parameters using these expressions,
and to compare with the existing bounds on the $S$, $T$ and $U$ parameters.

We presented this analysis for the operator $O_{tW}$ in Ref.~\cite{Greiner:2011tt}.
We found that the $S$ parameter is divergent, which is not surprising because in an effective theory there are two dimension-six operators that contribute to the $S$ and $T$ parameters, respectively, at tree level:
\begin{eqnarray}
O_{WB}&=&(\phi^\dagger\tau^I\phi)W^I_{\mu\nu}B^{\mu\nu}\label{op10},\\
O_\phi^{(3)}&=&(\phi^\dagger D^\mu \phi)[(D_\mu\phi)^\dagger\phi].\label{op11}
\end{eqnarray}
The divergent terms in the $S$ and $T$ parameters
are absorbed by renormalization of the coefficients of these two operators,
so $S$ and $T$ do not give useful information on the size of $O_{tW}$.\footnote{
We found that, rather than being divergent, the contribution of $O_{tW}$ to the $T$ parameter
vanishes \cite{Greiner:2011tt}.
A top-quark model that gives a nonvanishing contribution to the $T$ parameter is discussed in Ref.~\cite{Pomarol:2008bh}.}
On the other hand, the $U$ parameter is finite, because there is no dimension-six operator that
contributes at tree level.
Using the experimental bound $U=0.06\pm0.10$ given in \cite{PDG},
we found
\begin{equation}\label{eq:Ubound}
\frac{C_{tW}}{\Lambda^2}=-0.7\pm 1.1\ {\rm TeV^{-2}}.
\end{equation}

\begin{figure}[hbt]
\centering
\includegraphics[scale=1]{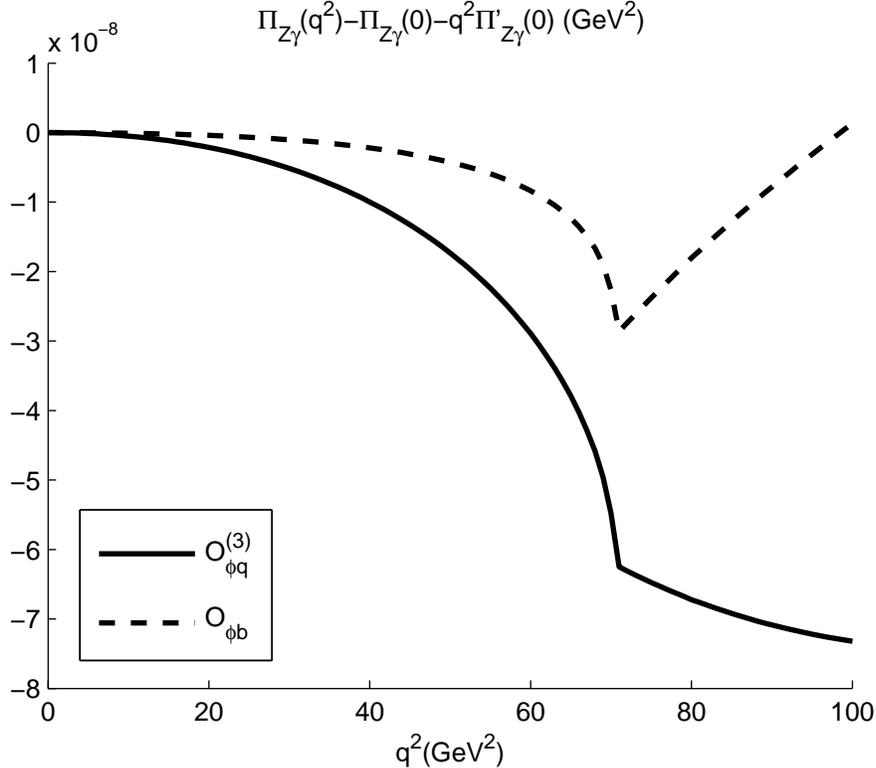}
\caption{The $q^2$ dependence of $\Pi_{Z\gamma}$.
The contributions from the operators $O_{\phi q}^{(3)}$ and $O_{\phi b}$
are shown for illustration.
A linear part in $q^2$ is subtracted so that $\Pi_{\gamma Z}(0)=\Pi_{\gamma Z}'(0)=0$.
There is a branch point at $q^2=4m_b^2$.\label{fig2}}
\end{figure}

Unfortunately, this kind of analysis is not appropriate for all nine operators.
The $S$, $T$, and $U$ parameters are defined by assuming a linear $q^2$ dependence of the self-energies.
However, once loop-level contributions are included,
the self-energies contain terms like $\ln q^2$ and $q^2\ln q^2$.
In particular, in a diagram with a bottom quark loop, the self-energies can have very different
$q^2$ dependence in the regions $q^2<4m_b^2$ and $q^2>4m_b^2$.
An example is shown in Figure \ref{fig2}.
Since the precision electroweak measurements
include data measured at both $q^2\approx0$ and $q^2\geq m_Z^2$, it is not reasonable to use a bound obtained by
assuming a linear $q^2$ dependence.
In addition, this calculation does not make full use of the obtained $q^2$ dependence of the self-energies.
By calculating the $U$ parameter, one can only put a constraint on one special linear combination of the
operators. The precision measurements, on the other hand, contain much more information.

In order to fully study the effects of the nine operators,
we will explicitly calculate the effect of self-energy corrections on all electroweak measurements,
and perform a fit including all operators in Eqs.~(\ref{op1})-(\ref{op9}) and Eqs.~(\ref{op10})-({\ref{op11}).
The self-energy corrections given in Appendix \ref{app:a} contain divergent terms.
Since all divergent terms are either constant or proportional to $q^2$,
they can contribute at most through the three oblique parameters.
Therefore all divergences can be properly absorbed once we include $O_{WB}$ and $O_\phi^{(3)}$ in our analysis. An effective field theory is renormalizable in this sense.

\section{Experiments}
\begin{table}[htb]\footnotesize
\begin{tabular}{|c|c|c|c|}
\hline
		&		Notation		&		Measurement		&		Reference
\\\hline
Z-pole	&	$\Gamma_Z$	& Total $Z$ width & \cite{PDG,:2005ema}
\\
				& $\sigma_{\rm had}$ &	Hadronic cross section &
\\
				&	$R_f$($f=e,\mu,\tau,b,c$)	& Ratios of decay rates &
\\
				& $A_{FB}^{0,f}$($f=e,\mu,\tau,b,c,s$) & Forward-backward asymmetries &
\\
				& $\bar{s}_l^2$ &	Hadronic charge asymmetry &
\\
				&	$A_f$($f=e,\mu,\tau,b,c,s$)		& Polarized asymmetries &
\\\hline
Fermion pair & $\sigma_f$($f=q,\mu,\tau,e)$ & Total cross sections for $e^+e^-\rightarrow f\bar{f}$
&\cite{Alcaraz:2006mx}
\\
production at LEP2 & $A_{FB}^f$($f=\mu,\tau$) & Forward-backward asymmetries for $e^+e^-\rightarrow f\bar{f}$ &\\\hline
$W$ mass & $m_W$ & $W$ mass from LEP and Tevatron & \cite{PDG}
\\
and decay rate& $\Gamma_W$ & $W$ width from Tevatron &
\\\hline
DIS & $Q_W(Cs)$	& Weak charge in Cs & \cite{PDG}
\\
and							& $Q_W(Tl)$ & Weak charge in Tl &
\\
atomic parity violation					& $Q_W(e)$	& Weak charge of the electron &
\\
																& $g_L^2,g_R^2$	 &$\nu_\mu$-nucleon scattering from NuTeV &
\\
																& $g_V^{\nu e},g_A^{\nu e}$& $\nu$-$e$ scattering from CHARM II &
\\\hline
$W$ pair production & $\sigma_W $  & Total cross section for $e^+e^-\rightarrow W^+W^-$ &\cite{Alcaraz:2006mx}
\\\hline
\end{tabular}
\caption{Major precision electroweak measurements used in this analysis. The total cross section for $e^+e^-\rightarrow e^+e^-$ is divergent. We use the cross section in the angular range $\cos\theta\in [-0.9,0.9]$ instead.\label{table}}
\end{table}
The measurements we use to constrain the coefficients of the operators are listed in Table \ref{table}.
Detailed descriptions for individual experiments can be found in the corresponding references.

For a given observable $X$, the prediction of the effective field theory can be written as
\begin{equation}
X_{\rm th}=X_{\rm SM}+\sum_i C_iX_i^{\rm dim 6},
\end{equation}
where $X_{\rm th}$ is the prediction in the presence of the operators,
$X_{\rm SM}$ is the Standard Model prediction, and $\sum_i C_iX_i^{\rm dim 6}$ are the corrections
from the new operators. Since only dimension-six operators are included,
higher-order terms in $C_i/\Lambda^2$ are dropped.

The SM predictions are computed to the required accuracy,
and can be found in the literature shown in Table \ref{table}.
The three most precisely measured electroweak observables,
$\alpha$, $G_F$, and $m_Z$, are taken to be the input parameters,
from which the SM gauge couplings and the Higgs VEV are inferred.
In addition, the following input parameters are used:
\begin{equation}
m_{\rm Higgs}=90^{+27}_{-22}\ {\rm GeV},
\quad m_t=173.2\pm 1.3\ {\rm GeV},
\quad \alpha_s(m_Z)=0.1183\pm0.0015,
\end{equation}
except for LEP2.
The sensitivities of the SM predictions to the input parameters for the fermion pair production
and $W$ pair production cross sections at LEP2 are negligible compared
to the experimental errors \cite{Han:2004az}. Therefore, we use the SM prediction
given in the corresponding references.

The corrections from the new operators include the tree-level contribution
to the self-energies from $O_{WB}$ and $O^{(3)}_{\phi}$,
the tree-level correction to the $Zb\bar{b}$ couplings from $O^{(3)}_{\phi q}$, $O^{(1)}_{\phi q}$ and
$O_{\phi b}$, and the loop-level contribution from all nine operators
in Eqs.~(\ref{op1})-(\ref{op9}) to the self-energies.
Once the self-energies are given, the corrections $X_i^{\rm dim 6}$ to all the experiments
can be obtained from the modified tree-level formulae for each observable. This will be discussed in the next section.

Given these results, we can calculate the total $\chi^2$ as a function of $C_i$:
\begin{equation}
\chi^2=\sum_X\frac{(X_{\rm th}-X_{\rm exp})^2}{\sigma_X^2}
=\sum_X\frac{(X_{\rm SM}-X_{\rm exp}+\sum_i C_iX_i^{\rm dim 6})^2}{\sigma_X^2},
\label{chi2}
\end{equation}
where $X_{\rm exp}$ is the experimental value for observable $X$ and $\sigma_X$ is the total error
which consists of both experimental and theoretical uncertainties.
The $\chi^2$ is a quadratic function of $C_i$.
The fit for the coefficients of the new operators is given by minimizing $\chi^2$.
The one-sigma bounds on the coefficients are given by $\chi^2-\chi_{\rm min}^2=1$.

Eq.~(\ref{chi2}) needs to be modified to account for the correlations between different measurements. There are two sets of data for which the correlations between measurements
cannot be neglected. These are the correlations between $Z$-pole observables \cite{:2005ema},
and the experimental error correlations for the hadronic total cross sections at LEP2 \cite{Alcaraz:2006mx}.
To include correlations, Eq.~(\ref{chi2}) should be modified to
\begin{equation}
\chi^2=\sum_{p,q}
(X_{\rm SM}^p-X_{\rm exp}^p+\sum_i C_iX_i^{p, \rm dim 6})
(\sigma^2)^{-1}_{pq}
(X_{\rm SM}^q-X_{\rm exp}^q+\sum_i C_iX_i^{q, \rm dim 6})
\end{equation}
where $X^{p,q}$ denotes different observables. The error matrix $\sigma^2$ is related
to the error $\sigma_p$ and the correlation matrix $\rho_{pq}$ by
\begin{equation}
\sigma^2_{pq}=\sigma_p\rho_{pq}\sigma_q
\end{equation}
The correlations for theoretical and experimental errors should be taken into account separately.

\section{Calculations}
In the presence of the new operators, the
corrections to the self-energies of $W$, $Z$ and $\gamma$ can be written as
\begin{equation}
\Pi_{XY}=\sum_iC_i{\Pi_{XY}}_i,
\end{equation}
where $\Pi_{XY}$ only includes the contributions from the new operators.
$(XY)=(ZZ)$,$(WW)$,\\$(\gamma\gamma)$,$(\gamma Z)$.

For the operators in Eqs.~(\ref{op1})-(\ref{op9}), the ${\Pi_{XY}}_i$'s are given in Appendix \ref{app:a}.
We also include $O_{WB}$ and $O_{\phi}^{(3)}$ in our calculation, so that
the divergences can be absorbed. For these two operators, the contributions at tree-level are:
\begin{eqnarray}
&&\Pi_{WW}=0
\label{pitree1},\\
&&\Pi_{ZZ}=C_{WB}\frac{2v^2}{\Lambda^2}s_Wc_Wq^2+C_{\phi}^{(3)}\frac{v^2}{2\Lambda^2}m_Z^2,\\
&&\Pi_{\gamma\gamma}=-C_{WB}\frac{2v^2}{\Lambda^2}s_Wc_Wq^2,\\
&&\Pi_{\gamma Z}=-C_{WB}\frac{v^2}{\Lambda^2}(c_W^2-s_W^2)q^2
\label{pitree2},
\end{eqnarray}
where $s_W=\sin\theta_W$ and $c_W=\cos\theta_W$.

In this section, we discuss the effect of self-energy corrections on each experiment.
We will show how to obtain the $C_iX_i^{\rm dim 6}$ term in Eq.~(\ref{chi2}).
We first illustrate the idea with an example.

For processes involving light fermions as external particles,
Peskin and Takeuchi have shown in Ref.~\cite{Peskin:1991sw} that
the corrections to the gauge boson self-energies can be incorporated
by a change in the couplings and gauge boson parameters.
For example, for electromagnetic interactions,
the coupling should be replaced by
\begin{equation}
\alpha_*(q^2)=\alpha_0(1+\Pi'_{\gamma\gamma}(q^2)),
\label{direct example}
\end{equation}
where $\alpha_0$ is the renormalized coupling, not including contributions from dimension-six operators.
The self energy $\Pi_{\gamma\gamma}(q^2)$ contains only the contributions from the dimension-six operators, and $\Pi'_{\gamma\gamma}(q^2)$ is defined as
\begin{equation}
\Pi'_{\gamma\gamma}(q^2)=\frac{\Pi_{\gamma\gamma}(q^2)}{q^2}.
\end{equation}

In the presence of a dimension-six operator, the renormalized coupling, $\alpha_0$,
is different from the coupling measured in experiments.
Therefore, the self-energy corrections affect the theoretical predictions in two different ways,
which we will call direct correction and indirect correction.
The direct correction is simply described by Eq.~(\ref{direct example}).
Any observable in an electromagnetic process is affected by a change in the coupling.
The indirect correction arises from the fact that we take the
fine structure constant as one of the input parameters.
The parameter $\alpha_0$ is then shifted from the measured fine structure constant $\alpha$,
which is measured at $q^2=0$. Thus, from Eq.~(\ref{direct example}),
\begin{equation}
\alpha=\alpha_0(1+\Pi'_{\gamma\gamma}(0)).
\label{indirect example}
\end{equation}
Therefore any observable that depends on the fine structure constant as an input parameter is affected by
Eq.~(\ref{indirect example}).
We can now eliminate $\alpha_0$ by combining Eqs.~(\ref{direct example}) and (\ref{indirect example}),
to obtain
\begin{flalign}
\alpha_*(q^2)= \alpha\left[1+\Pi'_{\gamma\gamma}(q^2)-\Pi'_{\gamma\gamma}(0)\right],
\label{alpha shift}
\end{flalign}
which can be used to calculate the correction to any electromagnetic observable.

We will show the direct correction and indirect correction to all observables in Section 4.1 and 4.2,
respectively, and combine them to calculate the total effects on all electroweak measurements,
except for the cross section for $W$ pair production.
The $W$ pair production cross section at LEP2 has relatively low statistics,
and thus we will only consider the tree-level contribution, {\itshape i.e.}~the contribution from
$O_{WB}$ and $O_{\phi}^{(3)}$.

\subsection{Direct correction}
In the SM, the matrix elements of the charged- and neutral-current interactions mediated by
electroweak gauge bosons can be written at tree level as
\begin{eqnarray}
&&\mathcal{M}_{\rm NC}=e^2\frac{QQ'}{q^2}+\frac{e^2}{s_W^2c_W^2}(I_3-s_W^2Q)\frac{1}{q^2-m_Z^2}(I'_3-s_W^2Q'),\\
&&\mathcal{M}_{\rm CC}=\frac{e^2}{2s_W^2}I_+\frac{1}{q^2-m_W^2}I_-.
\end{eqnarray}

Peskin and Takeuchi have shown in Ref.~\cite{Peskin:1991sw}
that the modification of the gauge boson self-energies
can be included by writing
\begin{eqnarray}
&&\mathcal{M}_{\rm NC}=e_*^2\frac{QQ'}{q^2}+\frac{e_*^2}{s_{W*}^2c_{W*}^2}(I_3-s_{W*}^2Q)\frac{Z_{Z*}}{q^2-m_{Z*}^2}(I'_3-s_{W*}^2Q')
\label{newamp1},\\
&&\mathcal{M}_{\rm CC}=\frac{e_*^2}{2s_{W*}^2}I_+\frac{Z_{W*}}{q^2-m_{W*}^2}I_-,
\label{newamp2}
\end{eqnarray}
where the starred quantities are functions of $q^2$:
\begin{eqnarray}
m_{W*}^2(q^2)&=&(1-Z_W)q^2+Z_W\left(m_{W0}^2+\Pi_{WW}(q^2)\right)\label{replace1},\\
m_{Z*}^2(q^2)&=&(1-Z_Z)q^2+Z_Z\left(m_{Z0}^2+\Pi_{ZZ}(q^2)\right),\\
Z_W&=&1+\frac{d}{dq^2}\Pi_{WW}(q^2)|_{q^2=m_W^2},\\
Z_Z&=&1+\frac{d}{dq^2}\Pi_{ZZ}(q^2)|_{q^2=m_Z^2},\\
Z_{W*}(q^2)&=&1+\frac{d}{dq^2}\Pi_{WW}(q^2)|_{q^2=m_W^2}-\Pi'_{\gamma\gamma}(q^2)-\frac{c_W}{s_W}\Pi'_{\gamma Z}(q^2),\\
Z_{Z*}(q^2)&=&1+\frac{d}{dq^2}\Pi_{ZZ}(q^2)|_{q^2=m_Z^2}-\Pi'_{\gamma\gamma}(q^2)-\frac{c_W^2-s_W^2}{s_Wc_W}\Pi'_{\gamma Z}(q^2),\\
s_{W*}^2(q^2)&=&s_{W0}^2-s_Wc_W\Pi'_{\gamma Z}(q^2),\label{sinw}\\
e_{*}^2(q^2)&=&e_0^2+e^2\Pi'_{\gamma\gamma}(q^2),\label{replace2}
\end{eqnarray}
where $\Pi'_{XY}(q^2)$ is defined as
\begin{equation}
\Pi'_{XY}(q^2)=\left(\Pi_{XY}(q^2)-\Pi_{XY}(0)\right)/q^2.
\end{equation}
The subscript 0 denotes the renormalized parameter, not including contributions from dimension-six operators.
When calculating the corrections due to dimension-six
operators, we can use tree-level relations between the renormalized parameters, such as
\begin{equation}\label{W0Z0}
m_{W0}^2=\frac{e_0^2}{s_{W0}^2}\frac{v^2}{4},\quad
m_{Z0}^2=\frac{e_0^2}{s_{W0}^2c_{W0}^2}\frac{v^2}{4}.
\end{equation}

Eqs.~(\ref{newamp1}) and (\ref{newamp2}) have exactly the same form as the tree-level SM amplitudes,
except that all the couplings and gauge-boson parameters are replaced by starred parameters.
This shows that the oblique corrections affect electroweak interaction observables only via the starred parameters.
In other words, given an observable in terms of renormalized parameters at tree-level,
we only need to replace the renormalized parameters with their starred counterparts
evaluated at
the appropriate momentum to incorporate the corrections from the self-energy diagrams.
For example, at tree-level the left-right asymmetry $A_e$ at the $Z$-pole is given by
\begin{equation}
A_e(m_Z^2)=\frac{2\left(1-4s_{W0}^2\right)}{1+\left(1-4s_{W0}^2\right)^2}.
\end{equation}
This is modified to
\begin{equation}
A_e(m_Z^2)=\frac{2\left(1-4s_{W*}^2(m_Z^2)\right)}{1+\left(1-4s_{W*}^2(m_Z^2)\right)^2}
\end{equation}
after the self-energy corrections are included.
Similarly, the $Z$ to $e^+e^-$ partial width is now corrected to
\begin{equation}
\Gamma_{e^+e^-}=\frac{e_*^2(m_Z^2)Z_{Z*}(m_Z^2)m_Z}{192\pi s_{W*}^2(m_Z^2)c_{W*}^2(m_Z^2)}
\left(\left(1-4s_{W*}^2(m_Z^2)\right)^2+1\right).
\end{equation}
Note that these corrections come from the difference between quantities with subscript $*$
and quantities with subscript $0$, therefore these are direct corrections.

For low energy measurements, it is more convenient to write
\begin{eqnarray}
&&\mathcal{M}_{\rm NC}=-4\sqrt{2}G_{F0}\left(1-\frac{1}{m_Z^2}\Pi_{ZZ}(0)\right)
\left(I_3-s_{W*}^2(0)Q\right)\left(I'_3-s_{W*}^2(0)Q'\right),
\label{newamp3}\\
&&\mathcal{M}_{\rm CC}=-2\sqrt{2}G_{F0}\left(1-\frac{1}{m_W^2}\Pi_{WW}(0)\right)I_+I_-,
\label{newamp4}
\end{eqnarray}
where
\begin{equation}
G_{F0}=\frac{1}{\sqrt{2}v^2}
\end{equation}
is the Fermi constant, not including contributions from dimension-six
operators. The direct corrections to any low energy observables
are thus incorporated by replacing $s_{W0}$ by $s_{W*}(0)$ and including an overall
factor of $(1-\Pi_{ZZ}(0)/m_Z^2)$ for neutral-current observables,
and $(1-\Pi_{WW}(0)/m_W^2)$ for charged-current observables.

\subsection{Indirect correction}
The indirect corrections arise from the shifts in the renormalized parameters.
The electroweak parameters ($g$, $g'$, $v$) are not directly measured.
Instead, we derive them from the most precisely measured observables
($\alpha$, $m_Z$, $G_F$).
When calculating the SM predictions for these observables,
the tree-level relations between ($g$, $g'$, $v$) and ($\alpha$, $m_Z$, $G_F$) are used.
When we include the new operators, the SM relations are altered.
This corresponds to a correction to all three input parameters.

To consider the indirect corrections,
we use ($\alpha_0$, $m_{Z0}$, $G_{F0}$) to denote the renormalized electroweak parameters, not including
contributions from dimension-six operators.
The relation between $\alpha$ and $\alpha_0$ can be read off from Eqs.~(\ref{newamp1}) and (\ref{replace2})
[see Eq.~(\ref{indirect example})]:
\begin{equation}
\alpha=\frac{e_*^2(0)}{4\pi}=\alpha_0(1+\Pi'_{\gamma\gamma}(0)).
\label{indirect1}
\end{equation}
The $Z$ mass $m_Z$ can be obtained by solving
$m_{Z*}^2(m_Z^2)=m_Z^2$; this gives
\begin{equation}
m_Z^2=m_{Z0}^2+\Pi_{ZZ}(m_Z^2).
\label{mz}
\end{equation}
The Fermi constant can be read off from Eq.~(\ref{newamp4}):
\begin{equation}
G_F=G_{F0}\left(1-\frac{1}{m_W^2}\Pi_{WW}(0)\right).
\end{equation}
Observables measured at energy scales above the $Z$ pole
are expressed in terms of ($\alpha$, $m_{Z}$, $s_W^2$) rather than ($\alpha$, $m_Z$, $G_F$).
To include indirect correction due to $s_W^2$, we need
\begin{eqnarray}
&&s_{W0}^2=\frac{1}{2}\left(1-\sqrt{1-\frac{4\pi\alpha_0}{\sqrt{2}G_{F0}m_{Z0}^2}}\right)
\nonumber\\
&&=\frac{1}{2}\left(1-\sqrt{1-\frac{4\pi\alpha}{\sqrt{2}G_{F}m_{Z}^2}}\right)
\left[
1-\frac{c_W^2}{c_W^2-s_W^2}\left(
\Pi'_{\gamma\gamma}(0)+\frac{1}{m_W^2}\Pi_{WW}(0)-\frac{1}{m_Z^2}\Pi_{ZZ}(m_Z^2)
\right)
\right]
\nonumber\\
&&=s_W^2
\left[
1-\frac{c_W^2}{c_W^2-s_W^2}\left(
\Pi'_{\gamma\gamma}(0)+\frac{1}{m_W^2}\Pi_{WW}(0)-\frac{1}{m_Z^2}\Pi_{ZZ}(m_Z^2)
\right)
\right].
\label{indirect2}
\end{eqnarray}
where
\begin{equation}
s_W^2=\frac{1}{2}\left(1-\sqrt{1-\frac{4\pi\alpha}{\sqrt{2}G_{F}m_{Z}^2}}\right)
\end{equation}
is the value for $s_W^2$ calculated at tree level using the observed values for ($\alpha$, $m_Z$, $G_F$).

Combining Eqs.~(\ref{replace1})-(\ref{replace2}) with Eqs.~(\ref{indirect1})-(\ref{indirect2})
to eliminate $\alpha_0$, $m_{Z0}$, $G_{F0}$, and $s_{W0}$,
we conclude that, for $q^2>0$, the total effect of direct and indirect corrections can be
incorporated by making the following replacement to the renormalized parameters
in the tree-level expressions for any observable:
\begin{eqnarray}
\alpha&\rightarrow&\alpha_*=\alpha+\delta\alpha
=\alpha\left(1+\Pi'_{\gamma\gamma}(q^2)-\Pi'_{\gamma\gamma}(0)\right)
\times\left\{
\begin{array}{ll}
1&\mbox{for interactions mediated by photon}\\
Z_{Z*}(q^2)&\mbox{for interactions mediated by $Z$ boson}\\
Z_{W*}(q^2)&\mbox{for interactions mediated by $W$ boson}
\end{array}
\right.,
\label{replace3}\\
m_Z^2&\rightarrow&m_{Z*}^2=m_Z^2+\delta m_Z^2
=m_Z^2-\Pi_{ZZ}(m_Z^2)+\Pi_{ZZ}(q^2)-(q^2-m_Z^2)\frac{d}{dq^2}\Pi_{ZZ}(q^2)|_{q^2=m_Z^2},\label{replace35}
\\
s_{W}^2&\rightarrow&s_{W*}^2=s_{W}^2+\delta s_{W}^2
=s_W^2\left[
1-\frac{c_W}{s_W}\Pi'_{\gamma Z}(q^2)-\frac{c_W^2}{c_W^2-s_W^2}\left(
\Pi'_{\gamma\gamma}(0)+\frac{1}{m_W^2}\Pi_{WW}(0)-\frac{1}{m_Z^2}\Pi_{ZZ}(m_Z^2)
\right)
\right].\label{replace4}
\end{eqnarray}
The renormalized parameters (subscript 0) are now completely eliminated,
and served only as intermediate quantities in the derivation.

For any observable measured at the $Z$-pole or above,
we can write it at tree level in terms of $\alpha$, $m_Z^2$ and $s_{W}^2$:
\begin{equation}
X_{\rm th}^{\rm tree}=X_{\rm th}^{\rm tree}(\alpha,m_Z^2,s_{W}^2).
\label{thtree}
\end{equation}
Therefore the contribution from the self-energy corrections can be written as
\begin{equation}
\delta X=C_iX_i^{\rm dim 6}=\frac{\partial X_{\rm th}^{\rm tree}}{\partial\alpha}\delta\alpha
+\frac{\partial X_{\rm th}^{\rm tree}}{\partial m_Z^2}\delta m_Z^2
+\frac{\partial X_{\rm th}^{\rm tree}}{\partial s_{W}^2}\delta s_{W}^2.
\label{dim6}
\end{equation}
Note that Eq.~(\ref{thtree}) is a tree-level relation, and we will not use it to compute
the entire theoretical prediction.
Instead, we use Eq.~(\ref{dim6}) to find the corrections which arise from the new operators.
Since these are already small corrections, the tree-level calculation is enough.
We then add them to the SM predictions including radiative corrections,
which are provided in the references shown in Table \ref{table}.

If the observables depend on the $Zb\bar{b}$ couplings, we will need to add to the r.h.s of Eq.~(\ref{dim6}) the following terms:
\begin{equation}
-\frac{v^2}{2\Lambda^2}\left(C_{\phi q}^{(3)}+C_{\phi q}^{(1)}+C_{\phi b}\right)\frac{\partial X_{\rm th}^{\rm tree}}{\partial g_V^b}
-\frac{v^2}{2\Lambda^2}\left(C_{\phi q}^{(3)}+C_{\phi q}^{(1)}-C_{\phi b}\right)\frac{\partial X_{\rm th}^{\rm tree}}{\partial g_A^b}.
\end{equation}
This accounts for the tree-level correction to the $Zb\bar{b}$ couplings from $C^{(3)}_{\phi q}$, $C^{(1)}_{\phi q}$ and
$C_{\phi b}$, given in Eqs.~(\ref{eq:Zbb1}) and (\ref{eq:Zbb2}).

For low energy measurements, we can now write
\begin{eqnarray}
&&\mathcal{M}_{\rm NC}=-4\sqrt{2}G_{F}\rho_*(0)
\left(I_3-s_{W*}^2(0)Q\right)\left(I'_3-s_{W*}^2(0)Q'\right)
\label{newamp5},\\
&&\mathcal{M}_{\rm CC}=-2\sqrt{2}G_{F}I_+I_-,
\label{newamp6}
\end{eqnarray}
where
\begin{equation}
\rho_*(0)=1-\frac{1}{m_Z^2}\Pi_{ZZ}(0)+\frac{1}{m_W^2}\Pi_{WW}(0).
\label{rho}
\end{equation}
The results of DIS and atomic parity violation experiments are usually expressed in terms
of the effective couplings in the neutral-current interactions.
The corrections to these results can thus be obtained by replacing $s_W^2$ by
\begin{equation}
s_{W*}^2(0)=s_W^2\left[
1-\frac{c_W}{s_W}\Pi'_{\gamma Z}(0)-\frac{c_W^2}{c_W^2-s_W^2}\left(
\Pi'_{\gamma\gamma}(0)+\frac{1}{m_W^2}\Pi_{WW}(0)-\frac{1}{m_Z^2}\Pi_{ZZ}(m_Z^2)
\right)
\right]\label{replace5}
\end{equation}
and including an overall factor of $\rho_*(0)$ to the couplings.

\subsection{Observables}
Now we proceed to consider the correction to each observable.
We will give the tree-level expressions for each observable, and then
use Eq.~(\ref{dim6}) to find the corrections that arise from the new operators.

\subsubsection{$Z$-pole process}
The process $e^+e^-\rightarrow f\bar{f}$ was studied around the $Z$-pole at SLC and LEP1.
At tree-level, the measured cross sections and asymmetries can be derived
from two quantities: the partial width of $Z\rightarrow f\bar{f}$, $\Gamma_{ff}$,
and the polarized asymmetry $A_f$.
The expressions are
\begin{eqnarray}
&&\Gamma_{ff}=\frac{\alpha m_Z}{12 s_W^2c_W^2}\left({g_V^f}^2+{g_A^f}^2\right),\\
&&A_f=\frac{2g_V^fg_A^f}{{g_V^f}^2+{g_A^f}^2},
\end{eqnarray}
where the $Z$-fermion couplings $g_V^f$ and $g_A^f$ are given by
\begin{equation}
\begin{array}{ccc}
f & g_V^f & g_A^f \\
\nu_e,\nu_\mu,\nu_\tau & +\frac{1}{2} & +\frac{1}{2}\\
e,\mu,\tau & -\frac{1}{2}+2s_W^2 & -\frac{1}{2}\\
u,c,t & +\frac{1}{2}-\frac{4}{3}s_W^2 & +\frac{1}{2}\\
d,s,b & -\frac{1}{2}+\frac{2}{3}s_W^2 & -\frac{1}{2}
\end{array}
\end{equation}
The $Z$-pole observables include:
\begin{itemize}
\item{Total width}
\begin{equation}
\Gamma_Z=\sum_f\Gamma_{ff}.
\end{equation}
\item{Total hadronic cross-section}
\begin{equation}
\sigma_h^0=\frac{12\pi}{m_Z^2}\frac{\Gamma_{ee}\Gamma_{\rm had}}{\Gamma_Z^2}.
\end{equation}
\item{Ratios of decay rates}
\begin{equation}
R_f=\left\{\begin{array}{ll}
\frac{\Gamma_{\rm had}}{\Gamma_{ff}} &\mbox{for $f=e,\mu,\tau$}\\
\frac{\Gamma_{ff}}{\Gamma_{\rm had}}&\mbox{for $f=b,c$}
\end{array}\right..
\end{equation}
\item{Forward-backward asymmetries}
\begin{equation}
A^{0,f}_{FB}=\frac{3}{4}A_eA_f,\quad f=e,\mu,\tau,b,c,s.
\end{equation}
\item{Effective angle extracted from the hadronic charge
asymmetry at LEP and from the combined lepton asymmetry from CDF and D0}
\begin{equation}
\bar{s}_l^2=s_W^2.
\end{equation}
\item{Polarized asymmetries}
\begin{equation}
A_f,\quad f=e,\mu,\tau,b,c,s.
\end{equation}
\end{itemize}
With these tree-level expressions,
we can apply Eq.~(\ref{dim6}) to derive the correction from the new operators.
For example, for the ratio $R_b$, we find
\begin{eqnarray}
\delta R_b
\!\!\!\!&=\!\!\!\!&
-24\frac{16s_W^4-36s_W^2+9}{(88s_W^4-84s_W^2+45)^2}
\left[
\frac{c_W}{s_W}\Pi'_{\gamma Z}(m_Z^2)+\frac{c_W^2}{c_W^2-s_W^2}\left(
\Pi'_{\gamma\gamma}(0)+\frac{1}{m_W^2}\Pi_{WW}(0)-\frac{1}{m_Z^2}\Pi_{ZZ}(m_Z^2)
\right)
\right]\nonumber\\
&&-24\frac{v^2}{\Lambda^2}\left(C_{\phi q}^{(3)}+C_{\phi q}^{(1)}\right)
\frac{40s_W^6-96s_W^4+72s_W^2-27}{(88s_W^4-84s_W^2+45)^2}
-48\frac{v^2}{\Lambda^2}C_{\phi b}
\frac{20s_W^4-18s_W^2+9}{(88s_W^4-84s_W^2+45)^2}.
\end{eqnarray}
Using the expressions for the $\Pi_{XY}$'s
given in Appendix \ref{app:a}, this can be written in the form of $C_iX_i^{\rm dim 6}$.

In practice, instead of deriving the above equation for all observables, we actually did the following:
for any given operator, we first give a small value to its coefficient $C_i$, then numerically compute
all starred quantities using Eqs.~(\ref{replace3}, \ref{replace35}, \ref{replace4}) and (\ref{rho}, \ref{replace5}).
We then take the tree-level expressions for all observables and substitute in the starred quantities, to obtain
the tree-level predictions for all these observables, including contributions from dimension-six operators.
Finally, we compare them with the SM tree-level predictions, to identify the corrections from dimension-six operators.

\subsubsection{Fermion pair production at LEP2}
The observables are the total cross-sections and forward-backward asymmetries for fermion pair production,
measured at different center of mass energies.
The matrix element for $e^+e^-\rightarrow f\bar{f}\ (f\neq e)$ is given by
\begin{eqnarray}
\mathcal{M}&=&\frac{4\pi\alpha}{(p+p')^2-m_Z^2+i\Gamma_Zm_Z}\frac{1}{4c_W^2s_W^2}
\bar{v}(p')\gamma^\mu\left(g_V^e-g_A^e\gamma^5\right)u(p)
\bar{u}(k)\gamma_\mu\left(g_V^f-g_A^f\gamma^5\right)v(k')\nonumber\\
&&-\frac{4\pi\alpha Q}{(p+p')^2}
\bar{v}(p')\gamma^\mu u(p)
\bar{u}(k)\gamma_\mu v(k'),
\end{eqnarray}
where $p$, $p'$ are the momenta of the incoming $e^+e^-$,
and $k$, $k'$ are the momenta of the outgoing fermions.
The cross-sections and forward-backward asymmetries can be calculated
from $\mathcal{M}$, and Eq.~(\ref{dim6}) can be applied
to obtain the corrections from the operators.

For $f=e$, there are additional contributions from the $t$-channel diagrams.
The matrix element is
\begin{eqnarray}
\mathcal{M}&=&\frac{4\pi\alpha}{(p+p')^2-m_Z^2+i\Gamma_Zm_Z}\frac{1}{4c_W^2s_W^2}
\bar{v}(p')\gamma^\mu\left(g_V^e-g_A^e\gamma^5\right)u(p)
\bar{u}(k)\gamma_\mu\left(g_V^e-g_A^e\gamma^5\right)v(k')\nonumber\\
&&-\frac{4\pi\alpha}{(p-k)^2-m_Z^2+i\Gamma_Zm_Z}\frac{1}{4c_W^2s_W^2}
\bar{u}(k)\gamma^\mu\left(g_V^e-g_A^e\gamma^5\right)u(p)
\bar{v}(p')\gamma_\mu\left(g_V^e-g_A^e\gamma^5\right)v(k')\nonumber\\
&&+\frac{4\pi\alpha}{(p+p')^2}
\bar{v}(p')\gamma^\mu u(p)
\bar{u}(k)\gamma_\mu v(k')
-\frac{4\pi\alpha}{(p-k)^2}
\bar{u}(k)\gamma^\mu u(p)
\bar{v}(p')\gamma_\mu v(k')
.
\end{eqnarray}

\subsubsection{$W$ mass}
The $W$ mass is measured both at the Tevatron and LEP2.
For the tree-level expression of $m_W$, we first solve $m_{W*}(m_W^2)=m_W^2$, which gives
\begin{equation}
m_W^2=m_{W0}^2+\Pi_{WW}(m_W^2).
\label{mw}
\end{equation}
Combining Eqs.~(\ref{W0Z0}), (\ref{mz}) and (\ref{indirect2}) with Eq.~(\ref{mw}),
we find that the correction to the $W$ mass is
\begin{equation}
\delta m_W^2=\Pi_{WW}(m_W^2)+\frac{s_W^2}{c_W^2-s_W^2}\Pi_{WW}(0)-\frac{c_W^4}{c_W^2-s_W^2}\Pi_{ZZ}(m_Z^2)
+\frac{s_W^2c_W^2}{c_W^2-s_w^2}m_Z^2\Pi'_{\gamma\gamma}(0).
\end{equation}

The $W$ width is measured at the Tevatron. The tree level expression is
\begin{equation}
\Gamma_W=\frac{3\alpha m_W}{4s_W^2}.
\end{equation}
The correction can be calculated using Eq.~(\ref{dim6}).

\subsubsection{DIS and atomic parity violation}
These are experiments performed at $q^2\approx0$.
These low energy observables are usually expressed in terms of the effective couplings $g_V^f$ and $g_A^f$,
which depend on $s_W^2$.
For the tree-level expressions, we will also include the factor $\rho_*(0)$,
which is 1 in the SM, and takes the value of Eq.~(\ref{rho}) in the presence of new operators:
\begin{equation}
\rho_*(0)=1+\delta\rho(0)=1-\frac{1}{m_Z^2}\Pi_{ZZ}(0)+\frac{1}{m_W^2}\Pi_{WW}(0).
\end{equation}
The correction to an observable $X$ is then given by
\begin{equation}
\delta X=C_iX_i^{\rm dim 6}=
\left.\frac{\partial X_{\rm th}^{\rm tree}}{\partial s_W^2}\right|_{\rho_*(0)=1}\delta s_W^2(0)
+\left.\frac{\partial X_{\rm th}^{\rm tree}}{\partial \rho_*(0)}\right|_{\rho_*(0)=1}\delta \rho(0).
\end{equation}

The observables include:
\begin{itemize}
\item{The weak charges for Cs and Tl, measured in atomic parity violation experiments.
The weak charge is given by}
\begin{equation}
Q_W(Z,N)=-2\left[(2Z+N)C_{1u}+(Z+2N)C_{1d}\right],
\end{equation}
where $Z$ and $N$ are the proton number and the neutron number of the atom.
The tree-level expressions for $C_{1u}$ and $C_{1d}$ are
\begin{equation}
C_{1u}=2\rho_*(0)g_A^eg_V^u,\quad
C_{1d}=2\rho_*(0)g_A^eg_V^d.
\end{equation}
\item{The weak charge of the electron, $Q_W(e)$ , measured in polarized M{\o}ller scattering:}
\begin{equation}
Q_W(e)=-2C_{2e}=-4\rho_*(0)g_A^eg_V^e.
\end{equation}
\item{The effective couplings $g_L$ and $g_R$ for $\nu$-nucleon scattering, measured at NuTeV. These are defined as}
\begin{equation}
g_L^2={g_{L,{\rm eff}}^{u2}}+{g_{L,{\rm eff}}^{d2}},\quad
g_R^2={g_{R,{\rm eff}}^{u2}}+{g_{R,{\rm eff}}^{d2}}.
\end{equation}
where $g_{L,{\rm eff}}^u$, $g_{R,{\rm eff}}^u$, $g_{L,{\rm eff}}^d$ and $g_{R,{\rm eff}}^d$
are the effective couplings between the $Z$ boson and the up and down quarks.
The tree-level expressions are
\begin{eqnarray}
g_{L,{\rm eff}}^u=\rho_*(0)\frac{g_V^u+g_A^u}{2},\quad
g_{L,{\rm eff}}^d=\rho_*(0)\frac{g_V^d+g_A^d}{2},\\
g_{R,{\rm eff}}^u=\rho_*(0)\frac{g_V^u-g_A^u}{2},\quad
g_{R,{\rm eff}}^d=\rho_*(0)\frac{g_V^d-g_A^d}{2}.\\
\end{eqnarray}
\item{The effective couplings $g_V^{\nu e}$ and $g_A^{\nu e}$ for $\nu$-e scattering,
measured at CHARM II. The expressions are}
\begin{equation}
g_V^{\nu e}=\rho_*(0)g_V^e,\quad g_A^{\nu e}=\rho_*(0)g_A^e.
\end{equation}
\end{itemize}

\subsubsection{$W$ pair production}
So far we have been using the approach of Peskin and Takeuchi
to study the effects of new operators. However, this approach only applies
to processes involving light fermions as external particles,
and cannot be used to study $W$ pair production.
Due to the relatively low statistics in the measurements, the constraints from $W$ pair production
are weak compared to other electroweak observables.
Therefore we will ignore all loop effects, and only focus on the effects of operators
$O_{WB}$ and $O_\phi^{(3)}$.

Using Eqs.~(\ref{pitree1})-(\ref{pitree2}) and Eq.~(\ref{indirect2}), we have
\begin{equation}
s_{W0}^2=
\frac{1}{2}\left(1-\sqrt{1-\frac{4\pi\alpha}{\sqrt{2}G_{F}m_{Z}^2}}\right)
\left[
1+\frac{c_W^2}{c_W^2-s_W^2}\left(
4C_{WB}\frac{v^2}{\Lambda^2}s_Wc_W+\frac{1}{2}C_{\phi}^{(3)}\frac{v^2}{\Lambda^2}
\right)
\right].
\end{equation}
Note that Eq.~(\ref{indirect2}) only has to do with the indirect corrections, so it is still valid.

The operator $O_{WB}$ changes the mixing of the $W^3$ and $B$ bosons.
We then define $s_{W*}$ and $c_{W*}$ as the new mixing angle,
\begin{equation}
s_{W*}=s_{W0}-C_{WB}\frac{v^2}{\Lambda^2}s_W^2c_W,\quad
c_{W*}=\sqrt{1-s_{W*}^2},
\end{equation}
so that the $W^+W^-Z$ and $W^+W^-\gamma $ vertices from the kinetic term $-\frac{1}{4}W_{\mu\nu}^IW^{I\mu\nu}$
have the same form as in the SM. Note that this definition of $s_{W*}$ is different from the one in Eq.~(\ref{sinw}),
which was only valid for light fermions.
In this way, the operators $O_{WB}$ and $O_\phi^{(3)}$
have the following effects:
\begin{itemize}
\item{The SM $Zf\bar{f}$ vertices are modified to:}
\begin{eqnarray}
\mathcal{L}_{Zf\bar{f}}=\frac{e}{s_{W*}c_{W*}}\left(1+C_{WB}\frac{v^2}{\Lambda^2}\frac{s_W}{c_W}\right)
Z_\mu\bar{f}\gamma^\mu\left(T^3P_L-s_{W*}^2\left(1+C_{WB}\frac{v^2}{\Lambda^2}\frac{c_{W}}{s_{W}}\right)Q_f\right)f,
\end{eqnarray}
where $P_L=(1-\gamma^5)/2$.
\item{The SM $W^+W^-Z$ and $W^+W^-\gamma $ vertices are modified. The contribution comes from $O_{WB}$:}
\begin{equation}
O_{WB}\rightarrow
-igC_{WB}\frac{v^2}{\Lambda^2}c_WA^{\mu\nu}W_\mu^+W_\nu^-
+igC_{WB}\frac{v^2}{\Lambda^2}s_WZ^{\mu\nu}W_\mu^+W_\nu^-.
\end{equation}
\item{The $W$ mass is changed to:}
\begin{equation}
m_W^2=m_Z^2\left(\frac{1}{2}\left(1+\sqrt{1-\frac{4\pi\alpha}{\sqrt{2}G_{F}m_{Z}^2}}\right)
-2C_{WB}\frac{v^2}{\Lambda^2}\frac{s_Wc_W^3}{c_W^2-s_W^2}-\frac{1}{2}C_\phi^{(3)}\frac{v^2}{\Lambda^2}\frac{c_W^4}{c_W^2-s_W^2}
\right).
\end{equation}
\end{itemize}
Using these results we can write down the matrix element. The process
has a $t$-channel contribution $\mathcal{M}_t$
and $s$-channel contributions $\mathcal{M}_{\gamma}$ and $\mathcal{M}_Z$,
which come from photon and $Z$ boson exchange. They are given by
\begin{eqnarray}
\mathcal{M}_t&=&-i\frac{e^2}{2s_{W*}^2}\bar{v}(p')\gamma^\mu\frac{1}{/\!\!\!k-/\!\!\!p'}
\gamma^\nu P_Lu(p)\epsilon_\mu^{*\lambda_1}\epsilon_\nu^{*\lambda_2}
,\\
\mathcal{M}_{\gamma}&=&
-ie^2\bar{v}(p')\gamma_\rho u(p)\frac{1}{q^2}\times\nonumber\\&&
\left[
\left(g^{\mu\nu}(k'-k)^\rho-g^{\nu\rho}(q+k')^\mu+g^{\mu\rho}(q+k)^\nu\right)
+C_{WB}\frac{v^2}{\Lambda^2}\frac{c_{W}}{s_{W}}
\left(g^{\mu\rho}q^\nu-g^{\nu\rho}q^\mu\right)
\right]
\epsilon_\mu^{*\lambda_1}\epsilon_\nu^{*\lambda_2}
,\\
\mathcal{M}_Z&=&
-i\frac{e^2}{s_{W*}^2}\left(1+C_{WB}\frac{v^2}{\Lambda^2}\frac{s_{W}}{c_{W}}\right)
\bar{v}(p')\gamma_\rho
\left[\frac{1}{2}P_L-\left(1+C_{WB}\frac{v^2}{\Lambda^2}\frac{c_{W}}{s_{W}}\right)s_{W*}^2\right]
u(p)
\frac{1}{q^2-m_Z^2}\times\nonumber\\&&
\left[
\left(g^{\mu\nu}(k'-k)^\rho-g^{\nu\rho}(q+k')^\mu+g^{\mu\rho}(q+k)^\nu\right)
-C_{WB}\frac{v^2}{\Lambda^2}\frac{s_{W}}{c_{W}}
\left(g^{\mu\rho}q^\nu-g^{\nu\rho}q^\mu\right)
\right]
\epsilon_\mu^{*\lambda_1}\epsilon_\nu^{*\lambda_2}.
\end{eqnarray}
where $p$, $p'$ are the momenta of the incoming $e^+e^-$,
$q=p+p'$,
$k$, $k'$ are the momenta of the outgoing $W^+W^-$, and
$\epsilon_\mu^{*\lambda_1}$, $\epsilon_\nu^{*\lambda_2}$
are the polarization vectors of the $W^+W^-$.
$q=p+p'$.
The cross section can be calculated from the matrix element.
The correction due to $O_{WB}$ and $O_\phi^{(3)}$
is obtained by taking the linear part in $C_{WB}$ and $C_\phi^{(3)}$.
Higher order terms in $C/\Lambda^2$ are neglected.

\subsection{Total $\chi^2$}
In the calculation of $\chi^2$, we choose the $\overline{\rm MS}$ scheme,
with the renormalization scale $\mu^2=m_Z^2$, even for processes in which
the characteristic scale is not $m_Z^2$.
This is done for ease of calculation. The error introduced by this choice
is of higher order and negligible.

We find that the contribution from the operator $O_{\phi\phi}$
is suppressed by the bottom quark mass, as can be seen from Eq.~(\ref{phiphi}).
Therefore we neglect this operator.
The contributions from operators $O_{bW}$ and $O_{bB}$
also have a factor of $m_b$. However, their effects can still be large,
because the expressions contain the function $b_0(m_b^2,m_b^2,q^2)$,
and its derivative with respect to $q^2$ is inversely proportional to $m_b^2$:
\begin{equation}
\left.\frac{d}{dq^2}b_0(m_b^2,m_b^2,q^2)\right|_{q^2=0}=-\frac{1}{6m_b^2}.
\end{equation}
Therefore, we will consider 10 operators:
\begin{equation}
O_{WB},\
O_\phi^{(3)},\
O_{\phi q}^{(3)},\
O_{\phi q}^{(1)},\
O_{\phi t},\
O_{\phi b},\
O_{tW},\
O_{bW},\
O_{tB},\
O_{bB}.
\end{equation}

Using Eq.~(\ref{chi2}), $\chi^2$ can be written as a quadratic function of $C_i$:
\begin{equation}
\chi^2=\chi_{\rm min}^2+(C_i-\hat{C}_i)M_{ij}(C_j-\hat{C}_j).
\label{chi22}
\end{equation}
Here $\chi_{\rm min}^2$ is the minimum $\chi^2$ in the presence of the new operators.
$\hat{C}_i$ corresponds to the best fit value for $C_i$.

In our calculation, we used the following input parameters:
\begin{eqnarray}
&\alpha(m_Z^2)=1/128.91,\qquad
G_F=1.166364\times10^{-5}\ \mathrm{GeV}^{-2},\qquad
m_Z=91.1876\ \mathrm{GeV},\nonumber\\
&m_t=172.9\ \mathrm{GeV},\qquad
m_b=4.79\ \mathrm{GeV}.
\end{eqnarray}
We find $\chi_{\rm min}^2=80.04$, the $\chi^2_{\rm min}$ per degree of freedom is 0.80,
compared with the SM value 0.84.
The matrix $M_{ij}$ and the best fit values $\hat{C}_i$ are given in Appendix \ref{app:b}.

\subsection{Global fit}
The one-sigma bounds on the operators are given by $\chi^2-\chi_{\rm min}^2=1$.
By diagonalizing the matrix $M_{ij}$, we find ten
linear combinations of $C_i$ that are statistically independent.
Their best fit values and one-sigma bounds are given by:
\begin{eqnarray}
\left(
\begin{array}{cccccccccc}
-0.961 & -0.273 & +0.029 & -0.004 & +0.024 & +0.000 & +0.012 & -0.000 & +0.015 & +0.001 \\
-0.064 & +0.159 & -0.701 & -0.680 & -0.015 & +0.130 & +0.001 & -0.000 & +0.001 & +0.000 \\
+0.268 & -0.940 & -0.063 & -0.182 & +0.088 & +0.002 & -0.022 & +0.002 & -0.005 & -0.001 \\
-0.008 & +0.019 & -0.095 & -0.086 & -0.003 & -0.991 & -0.019 & +0.001 & -0.000 & -0.000 \\
-0.014 & -0.065 & -0.336 & +0.344 & -0.389 & +0.017 & -0.768 & +0.057 & -0.135 & -0.039 \\
+0.009 & -0.107 & -0.326 & +0.322 & -0.590 & -0.009 & +0.623 & -0.065 & -0.194 & +0.014 \\
-0.016 & +0.030 & +0.137 & -0.137 & +0.128 & -0.000 & -0.003 & +0.138 & -0.935 & -0.229 \\
-0.004 & +0.008 & +0.034 & -0.034 & +0.039 & +0.001 & -0.094 & -0.745 & -0.244 & +0.610 \\
-0.001 & +0.001 & -0.003 & +0.003 & +0.007 & -0.000 & +0.025 & +0.646 & -0.090 & +0.757 \\
-0.001 & +0.001 & +0.505 & -0.505 & -0.689 & -0.000 & -0.108 & +0.014 & +0.054 & +0.009
\end{array}
\right)
\nonumber\\
\times
\frac{1}{\Lambda^2}
\left(
\begin{array}{c}
C_{WB}\\C_\phi^{(3)}\\C_{\phi q}^{(3)}\\C_{\phi q}^{(1)}\\
C_{\phi t}\\C_{\phi b}\\C_{tW}\\C_{bW}\\C_{tB}\\C_{bB}
\end{array}
\right)
=
\left(
\begin{array}{rl}
-0.0004 &\pm 0.0029 \\
-0.013 &\pm 0.014 \\
+0.011 &\pm  0.023 \\
 +0.59 &\pm  0.27 \\
 -0.22 &\pm    1.10 \\
  -1.76 &\pm   1.63 \\
  -2.2 &\pm   11.9 \\
  -9.2 &\pm   21.1 \\
   +102.4 &\pm   50.4 \\
-1.36\mbox{e+3} &\pm 1.38\mbox{e+3}
\end{array}
\right)
{\rm TeV}^{-2}.
\end{eqnarray}
where the $10\times10$ matrix is orthogonal.
We can see that in the first row and the third row,
the first two components are much larger than the other components.
This means that these two rows approximately correspond to
constraints on the coefficients $C_{WB}$ and $C_\phi^{(3)}$,
or equivalently, the $S$ and $T$ parameters.
Since we are interested in the other eight operators,
we can assume that $C_{WB}$ and $C_\phi^{(3)}$ always take the values
that minimize the $\chi^2$. This doesn't mean that we fix $C_{WB}$ and $C_\phi^{(3)}$. The values that minimize the total $\chi^2$ depend on the values of the other eight coefficients. In other words, we let these two coefficients freely float.  In this way, we find the following constraints on the eight operators:
\begin{eqnarray}
\left(
\begin{array}{cccccccc}
-0.702 & -0.701 & -0.000 & +0.128 & -0.003 & +0.000 & -0.000 & -0.000 \\
-0.094 & -0.087 & -0.002 & -0.992 & -0.019 & +0.001 & -0.001 & -0.000 \\
-0.342 & +0.349 & -0.398 & +0.017 & -0.761 & +0.056 & -0.136 & -0.039 \\
-0.326 & +0.323 & -0.591 & -0.009 & +0.632 & -0.065 & -0.191 & +0.015 \\
+0.137 & -0.137 & +0.128 & -0.000 & -0.003 & +0.138 & -0.935 & -0.229 \\
+0.034 & -0.034 & +0.039 & +0.001 & -0.094 & -0.745 & -0.244 & +0.610 \\
-0.003 & +0.003 & +0.007 & -0.000 & +0.025 & +0.646 & -0.090 & +0.757 \\
+0.505 & -0.505 & -0.689 & -0.000 & -0.108 & +0.014 & +0.054 & +0.009
\end{array}
\right)
\nonumber\\\times
\frac{1}{\Lambda^2}
\left(
\begin{array}{c}
C_{\phi q}^{(3)}\\C_{\phi q}^{(1)}\\
C_{\phi t}\\C_{\phi b}\\C_{tW}\\C_{bW}\\C_{tB}\\C_{bB}
\end{array}
\right)
=
\left(
\begin{array}{rl}
-0.011 &\pm 0.014 \\
 +0.59 &\pm  0.27 \\
 -0.23 &\pm    1.10 \\
  -1.75 &\pm   1.62 \\
  -2.2 &\pm   11.9 \\
  -9.2 &\pm   21.1 \\
   +102.4 &\pm   50.4 \\
-1.36\mbox{e+3} &\pm 1.38\mbox{e+3} \\
\end{array}
\right)
{\rm TeV}^{-2}.\label{result}
\end{eqnarray}

This is the main result of this paper.
We can see that the first row is approximately a constraint on
$(O_{\phi q}^{(3)}+O_{\phi q}^{(1)})/\sqrt{2}$,
which corresponds to the left-handed $Zb\bar{b}$ coupling,
while the second row corresponds to a constraint on $O_{\phi b}$,
which is the right-handed $Zb\bar{b}$ coupling.
These are the tightest bounds, since the contribution arises at tree-level.
The other constraints are mainly from loop-level effects.
The third row is approximately a bound on $C_{tW}$.
This can be compared with the bound obtained from
the $U$ parameter, Eq.~(\ref{eq:Ubound}).
The results are in close agreement.
The remaining five rows yield bounds on linear combinations of operators.

\section{Conclusions}\label{sec:conclusions}
In this paper, we have studied the effects of non-standard top quark couplings
on precision electroweak measurements.
The top quark plays a role as a virtual particle in these measurements.
Our study is based on an effective field theory approach,
which allows us to calculate the self-energies of the electroweak gauge bosons at loop-level.

We have examined the effects of nine dimension-six operators that generate non-standard couplings
between electroweak gauge bosons and the third generation quarks.
These operators mainly contribute through loop corrections to the gauge boson self-energies,
but some of them also have a tree-level contribution to the $Zb\bar{b}$ couplings.
The contribution of one of the operators is suppressed
by $m_b$ and is hence negligible.
We have also included the operators $O_{WB}$ and $O_\phi^{(3)}$, in order to deal with the divergences
that appear in our calculation.

We have calculated the total $\chi^2$ and performed a global fit including these ten operators.
We allow $C_{WB}$ and $C_\phi^{(3)}$ to vary,
and thus obtain bounds on eight dimension-six operators.
The result is shown in Eq.~(\ref{result}).
The two tightest bounds are from tree-level contributions to the
$Zb\bar{b}$ couplings, $g_L^b$ and $g_R^b$, and the other bounds are from loop-level contributions.
The best bound from loop-level contribution constrains $\frac{C}{\Lambda^2}$
to be of order 1 $\mathrm{TeV}^{-2}$.

If the bound on some linear combination of $C_i$ is too weak, this bound cannot be trusted,
because the linear analysis is not applicable if the coefficients are large.
The contribution from dimension-six operators relative to the SM contribution
is of order $g^2/(4\pi)^2\times\frac{C_iv^2}{\Lambda^2}$.
This should be less than unity, thus
\begin{equation}
\frac{C_i}{\Lambda^2}\ll6.1\times 10^{3} {\rm\ TeV}^{-2}
\end{equation}
Even the weakest bound from our results does not exceed this limit,
therefore all bounds can be trusted.

Using Eq.~(\ref{chi22}), one can put constraints on a subset of these operators.
For example, the one-sigma bound on the coefficient $C_i$,
assuming only one coefficient deviates from its best fit value, is given by
$\hat{C_i}\pm M_{ii}^{-1/2}$, where $M_{ii}$ is the diagonal element of the matrix $M$
and is not summed over $i$.

We can also consider only one coefficient to be nonzero at a time.
In this case, we found the constraints on each individual coefficient are:
\begin{flalign}
\frac{C_{\phi q}^{(3)}}{\Lambda^2}+\frac{C_{\phi q}^{(1)}}{\Lambda^2}&=0.016\pm 0.021 {\rm\ TeV}^{-2}\\
\frac{C_{\phi q}^{(3)}}{\Lambda^2}-\frac{C_{\phi q}^{(1)}}{\Lambda^2}&=2.0\pm 2.7 {\rm\ TeV}^{-2}\\
\frac{C_{\phi t}}{\Lambda^2}&=1.8\pm 1.9 {\rm\ TeV}^{-2}\\
\frac{C_{\phi b}}{\Lambda^2}&=-0.16\pm 0.10 {\rm\ TeV}^{-2}\\
\frac{C_{tW}}{\Lambda^2}&=-0.4\pm 1.2  {\rm\ TeV}^{-2}\label{eq:boundtW}\\
\frac{C_{bW}}{\Lambda^2}&=11\pm 13 {\rm\ TeV}^{-2}\\
\frac{C_{tB}}{\Lambda^2}&=4.8\pm 5.3 {\rm\ TeV}^{-2}\\
\frac{C_{bB}}{\Lambda^2}&=8\pm 19 {\rm\ TeV}^{-2}
\end{flalign}

Some of these operators are already constrained from other measurements.
The operator $O_{tW}$ modifies the top-quark branching ratio to zero-helicity
$W$ bosons \cite{Scott}.
Recently, the combination of CDF and D0 measurements of the $W$-boson helicity in top-quark decays reports a measurement of $f_0$, the zero-helicity fraction \cite{Whelicity}:
\begin{equation}
f_0=0.685\pm 0.057 [\pm 0.035 (\rm{stat.})\pm 0.045(\rm{syst.})]
\end{equation}
This yields the constraint on $C_{tW}$:
\begin{equation}
\frac{C_{tW}}{\Lambda^2}=0.03\pm0.94\ {\rm TeV}^{-2}\label{eq:Tevbound}
\end{equation}
We see that Eqs.~(\ref{eq:Ubound}), (\ref{eq:boundtW}), and (\ref{eq:Tevbound})
give similar constraints.

Constraints may also be gleaned from $B$ physics.
The operators $O_{\phi q}^{(3)}-O_{\phi q}^{(1)}$ and $O_{\phi\phi}$
affect the branching ratio for $\bar{B}\to X_s\gamma$.
The following constraints are obtained in Ref.~\cite{Grzadkowski:2008mf}:
\begin{flalign}
\frac{C_{\phi q}^{(3)}}{\Lambda^2}-\frac{C_{\phi q}^{(1)}}{\Lambda^2}&=-1.6\pm 1.3 {\rm\ TeV}^{-2}\\
\frac{C_{\phi\phi}}{\Lambda^2}&=0.030\pm 0.026 {\rm\ TeV}^{-2}\,.
\end{flalign}
There are also contributions from $O_{tW}$ and $O_{bW}$, but these are ultraviolet divergent. Thus there must be a tree-level contribution from another dimension-six operator, which masks the contributions from $O_{tW}$ and $O_{bW}$. Therefore one cannot obtain bounds on these two operators.

In addition, constraints on the operator $O_{tW}$
can also be obtained from $B_{d,s}-\bar{B}_{d,s}$ mixing \cite{Drobnak:2011wj}:
\begin{flalign}
\frac{C_{tW}}{\Lambda^2}&=-0.06\pm 0.79 {\rm\ TeV}^{-2}\,.
\end{flalign}
This is comparable to the bounds given in Eqs.~(\ref{eq:Ubound}), (\ref{eq:boundtW}), and (\ref{eq:Tevbound}).
There is also a contribution from $O_{\phi q}^{(3)}-O_{\phi q}^{(1)}$
that is divergent and requires a tree-level contribution, so this operator cannot be bounded from
$B_{d,s}-\bar{B}_{d,s}$ mixing.

Finally, we summarize all these constraints in Table~\ref{table:constraints}.
\begin{table}[htb]
\centering
{\begin{tabular}{|c|c|c|c|c|}
\hline
Coefficients & Electroweak data & $W$ helicity & $\bar{B}\to X_s\gamma$ & $B_{d,s}-\bar{B}_{d,s}$ mixing \\\hline
$\left(C_{\phi{q}}^{(3)}+C_{\phi{q}}^{(1)}\right)/\Lambda^2$& $0.016 \pm  0.021 $&                  &        &        \\ \hline
$\left(C_{\phi{q}}^{(3)}-C_{\phi{q}}^{(1)}\right)/\Lambda^2$ & $2.0 \pm  2.7 $&   & $-1.6 \pm  1.3$ & $   $  \\ \hline
$C_{\phi{t}}/\Lambda^2$ & $1.8 \pm  1.9 $&                  &                 &                  \\ \hline
$C_{\phi{b}}/\Lambda^2$ & $-0.16 \pm  0.10$ &                  &        &                  \\ \hline
$C_{\phi\phi}/\Lambda^2$ &             &                  &     $0.030 \pm  0.026$  &         \\ \hline
$C_{tW}/\Lambda^2$ & $-0.4 \pm  1.2 $& $0.03 \pm  0.94 $&   &$ -0.06 \pm  0.79$ \\ \hline
$C_{bW}/\Lambda^2$ & $11 \pm  13$ &                 &     &                  \\ \hline
$C_{tB}/\Lambda^2$ & $4.8 \pm  5.3$ &                 &                 &                  \\ \hline
$C_{bB}/\Lambda^2$ & $8 \pm  19 $&                  &                  &                 \\\hline
\end{tabular}}
\caption{Bounds on operators, in units of ${\rm\ TeV}^{-2}$.\label{table:constraints}}
\end{table}

We can see that for most operators our analysis gives the best bounds available,
because electroweak data is the only way to access these operators so far.
For the other operators, our bounds are comparable to the best bounds obtained from colliders and $B$ physics.

\section{Acknowledgements}
We are grateful for conversations and correspondence with T.~Liss, J.~Serra, A.~Strumia, J.~Thaler, and J.~M.~Yang.
This material is based upon work supported in part by the U.~S. Department of Energy
under contracts No.~DE-FG02-91ER40677.

\vspace{1in}
\appendix
\section{Gauge boson self-energies}
\label{app:a}
Here we give $\Pi_{XY}$ for all 9 operators.
\begin{itemize}

\item{$O_{\phi q}^{(3)}$}
\begin{eqnarray}
\Pi_{WW}&=&-N_c\frac{g^2}{4\pi^2}\frac{v^2}{\Lambda^2}
\left[
\left(\frac{1}{6}q^2-\frac{1}{4}(m_t^2+m_b^2)\right)E
\right.\nonumber\\&&\left.
-q^2b_2(m_t^2,m_b^2,q^2)+\frac{1}{2}\left(m_b^2b_1(m_t^2,m_b^2,q^2)+m_t^2b_1(m_b^2,m_t^2,q^2)\right)
\right]
\\
\Pi_{ZZ}&=&-N_c\frac{g^2}{\cos^2\theta_W}\frac{1}{4\pi^2}\frac{v^2}{\Lambda^2}
\left[
\left(\frac{1}{6}(1-\sin^2\theta_W)q^2-\frac{1}{4}(m_t^2+m_b^2)\right)E
\right.\nonumber\\&&\left.
-q^2\left(\left(\frac{1}{2}-\frac{2}{3}\sin^2\theta_W\right)b_2(m_t^2,m_t^2,q^2)+\left(\frac{1}{2}-\frac{1}{3}\sin^2\theta_W\right)b_2(m_b^2,m_b^2,q^2)\right)
\right.\nonumber\\&&\left.
+\frac{1}{4}\left(m_t^2b_0(m_t^2,m_t^2,q^2)+m_b^2b_0(m_b^2,m_b^2,q^2)\right)
\right]
\\
\Pi_{\gamma\gamma}&=&0
\\
\Pi_{\gamma Z}&=&-N_cg^2\frac{\sin\theta_W}{\cos\theta_W}\frac{1}{8\pi^2}\frac{v^2}{\Lambda^2}
\left[\frac{1}{6}E-\frac{2}{3}b_2(m_t^2,m_t^2,q^2)-\frac{1}{3}b_2(m_b^2,m_b^2,q^2)\right]q^2
\end{eqnarray}

\item{$O_{\phi q}^{(1)}$}
\begin{eqnarray}
\Pi_{WW}&=&0
\\
\Pi_{ZZ}&=&N_c\frac{g^2}{\cos^2\theta_W}\frac{1}{4\pi^2}\frac{v^2}{\Lambda^2}
\left[-\left(\frac{1}{4}m_t^2-\frac{1}{4}m_b^2+\frac{1}{18}q^2\sin^2\theta_W\right)E
\right.\nonumber\\&&\left.
-q^2\left(\left(\frac{1}{2}-\frac{2}{3}\sin^2\theta_W\right)b_2(m_t^2,m_t^2,q^2)
-\left(\frac{1}{2}-\frac{1}{3}\sin^2\theta_W\right)b_2(m_b^2,m_b^2,q^2)\right)
\right.\nonumber\\&&\left.
+\frac{1}{4}\left(m_t^2b_0(m_t^2,m_t^2,q^2)-m_b^2b_0(m_b^2,m_b^2,q^2)\right)
\right]
\\
\Pi_{\gamma\gamma}&=&0
\\
\Pi_{\gamma Z}&=&N_cg^2\frac{\sin\theta_W}{\cos\theta_W}\frac{1}{8\pi^2}\frac{v^2}{\Lambda^2}
\left[\frac{1}{18}E-\frac{2}{3}b_2(m_t^2,m_t^2,q^2)+\frac{1}{3}b_2(m_b^2,m_b^2,q^2)\right]q^2
\end{eqnarray}

\item{$O_{\phi t}$}
\begin{eqnarray}
\Pi_{WW}&=&0
\\
\Pi_{ZZ}&=&N_c\frac{g^2}{\cos^2\theta_W}\frac{1}{4\pi^2}\frac{v^2}{\Lambda^2}
\left[\left(\frac{1}{4}m_t^2-\frac{1}{9}q^2\sin^2\theta_W\right)E
\right.\nonumber\\&&\left.
-\left(\frac{1}{4}m_t^2b_0(m_t^2,m_t^2,q^2)-\frac{2}{3}q^2\sin^2\theta_Wb_2(m_t^2,m_t^2,q^2)\right)
\right]
\\
\Pi_{\gamma\gamma}&=&0
\\
\Pi_{\gamma Z}&=&N_cg^2\frac{\sin\theta_W}{\cos\theta_W}\frac{1}{12\pi^2}\frac{v^2}{\Lambda^2}
\left(\frac{1}{6}E-b_2(m_t^2,m_t^2,q^2)\right)q^2
\end{eqnarray}

\item{$O_{\phi b}$}
\begin{eqnarray}
\Pi_{WW}&=&0
\\
\Pi_{ZZ}&=&N_c\frac{g^2}{\cos^2\theta_W}\frac{1}{4\pi^2}\frac{v^2}{\Lambda^2}
\left[-\left(\frac{1}{4}m_b^2-\frac{1}{18}q^2\sin^2\theta_W\right)E
\right.\nonumber\\&&\left.
+\left(\frac{1}{4}m_b^2b_0(m_b^2,m_b^2,q^2)-\frac{1}{3}q^2\sin^2\theta_Wb_2(m_b^2,m_b^2,q^2)\right)
\right]
\\
\Pi_{\gamma\gamma}&=&0
\\
\Pi_{\gamma Z}&=&-N_cg^2\frac{\sin\theta_W}{\cos\theta_W}\frac{1}{24\pi^2}\frac{v^2}{\Lambda^2}
\left(\frac{1}{6}E-b_2(m_b^2,m_b^2,q^2)\right)q^2
\end{eqnarray}

\item{$O_{\phi\phi}$}
\begin{eqnarray}
\Pi_{WW}&=&-N_cg^2\frac{1}{16\pi^2}\frac{v^2}{\Lambda^2}m_tm_b\left(E-b_0(m_t^2,m_b^2,q^2)\right)
\label{phiphi}
\\
\Pi_{ZZ}&=&0
\\
\Pi_{\gamma\gamma}&=&0
\\
\Pi_{\gamma Z}&=&0
\end{eqnarray}

\item{$O_{tW}$}
\begin{eqnarray}
\Pi_{WW}&=&-N_cg\frac{\sqrt{2}}{4\pi^2}\frac{vm_t}{\Lambda^2}\left(\frac{1}{2}E-b_1(m_b^2,m_t^2,q^2)\right)q^2
\\
\Pi_{ZZ}&=&-N_cg\frac{\sqrt{2}}{4\pi^2}\frac{vm_t}{\Lambda^2}\left(\frac{1}{2}-\frac{4}{3}\sin^2\theta_W\right)
\left(E-b_0(m_t^2,m_t^2,q^2)\right)q^2
\\
\Pi_{\gamma\gamma}&=&-N_cg\frac{\sqrt{2}}{4\pi^2}\frac{vm_t}{\Lambda^2}\frac{4}{3}\sin^2\theta_W\left(E-b_0(m_t^2,m_t^2,q^2)\right)q^2
\\
\Pi_{\gamma Z}&=&-N_cg\frac{\sqrt{2}}{4\pi^2}\frac{vm_t}{\Lambda^2}\frac{\sin\theta_W}{\cos\theta_W}
\left(\frac{11}{12}-\frac{4}{3}\sin^2\theta_W\right)\left(E-b_0(m_t^2,m_t^2,q^2)\right)q^2
\end{eqnarray}

\item{$O_{bW}$}
\begin{eqnarray}
\Pi_{WW}&=&-N_cg\frac{\sqrt{2}}{4\pi^2}\frac{vm_b}{\Lambda^2}\left(\frac{1}{2}E-b_1(m_t^2,m_b^2,q^2)\right)q^2
\\
\Pi_{ZZ}&=&-N_cg\frac{\sqrt{2}}{4\pi^2}\frac{vm_b}{\Lambda^2}\left(\frac{1}{2}-\frac{2}{3}\sin^2\theta_W\right)
\left(E-b_0(m_b^2,m_b^2,q^2)\right)q^2
\\
\Pi_{\gamma\gamma}&=&-N_cg\frac{\sqrt{2}}{4\pi^2}\frac{vm_b}{\Lambda^2}\frac{2}{3}\sin^2\theta_W\left(E-b_0(m_b^2,m_b^2,q^2)\right)q^2
\\
\Pi_{\gamma Z}&=&-N_cg\frac{\sqrt{2}}{4\pi^2}\frac{vm_b}{\Lambda^2}\frac{\sin\theta_W}{\cos\theta_W}
\left(\frac{7}{12}-\frac{2}{3}\sin^2\theta_W\right)\left(E-b_0(m_b^2,m_b^2,q^2)\right)q^2
\end{eqnarray}

\item{$O_{tB}$}
\begin{eqnarray}
\Pi_{WW}&=&0
\\
\Pi_{ZZ}&=&N_cg\frac{\sqrt{2}}{4\pi^2}\frac{vm_t}{\Lambda^2}\frac{\sin\theta_W}{\cos\theta_W}\left(\frac{1}{2}-\frac{4}{3}\sin^2\theta_W\right)
\left(E-b_0(m_t^2,m_t^2,q^2)\right)q^2
\\
\Pi_{\gamma\gamma}&=&-N_cg\frac{\sqrt{2}}{4\pi^2}\frac{vm_t}{\Lambda^2}\frac{4}{3}\sin\theta_W
\cos\theta_W\left(E-b_0(m_t^2,m_t^2,q^2)\right)q^2
\\
\Pi_{\gamma Z}&=&-N_cg\frac{\sqrt{2}}{4\pi^2}\frac{vm_t}{\Lambda^2}
\left(\frac{1}{4}-\frac{4}{3}\sin^2\theta_W\right)\left(E-b_0(m_t^2,m_t^2,q^2)\right)q^2
\end{eqnarray}

\item{$O_{bB}$}
\begin{eqnarray}
\Pi_{WW}&=&0
\\
\Pi_{ZZ}&=&-N_cg\frac{\sqrt{2}}{4\pi^2}\frac{vm_b}{\Lambda^2}\frac{\sin\theta_W}{\cos\theta_W}\left(\frac{1}{2}-\frac{2}{3}\sin^2\theta_W\right)
\left(E-b_0(m_b^2,m_b^2,q^2)\right)q^2
\\
\Pi_{\gamma\gamma}&=&N_cg\frac{\sqrt{2}}{4\pi^2}\frac{vm_b}{\Lambda^2}\frac{2}{3}\sin\theta_W
\cos\theta_W\left(E-b_0(m_b^2,m_b^2,q^2)\right)q^2
\\
\Pi_{\gamma Z}&=&N_cg\frac{\sqrt{2}}{4\pi^2}\frac{vm_b}{\Lambda^2}
\left(\frac{1}{4}-\frac{2}{3}\sin^2\theta_W\right)\left(E-b_0(m_b^2,m_b^2,q^2)\right)q^2
\end{eqnarray}
\end{itemize}

Here $\theta_W$ is the weak angle, $N_c=3$ is the number of colors.
$E=\frac{2}{4-d}-\gamma+\ln 4\pi
$, and the functions $b_i$ are given by
\begin{eqnarray}
&&b_0(m_1^2,m_2^2,q^2)=\int_0^1\ln\frac{(1-x)m_1^2+xm_2^2-x(1-x)q^2}{\mu^2}dx,\\
&&b_1(m_1^2,m_2^2,q^2)=\int_0^1x\ln\frac{(1-x)m_1^2+xm_2^2-x(1-x)q^2}{\mu^2}dx,\\
&&b_2(m_1^2,m_2^2,q^2)=\int_0^1x(1-x)\ln\frac{(1-x)m_1^2+xm_2^2-x(1-x)q^2}{\mu^2}dx,
\end{eqnarray}
where $\mu$ is the 't Hooft mass.
They have the following analytical expressions:
\begin{eqnarray}
b_0(m_1^2,m_2^2,q^2)&=&-2+\log\frac{m_1m_2}{\mu^2}+\frac{m_1^2-m_2^2}{q^2}\log\left(\frac{m_1}{m_2}\right)
\nonumber\\&&
+\frac{1}{q^2}\sqrt{|(m_1+m_2)^2-q^2||(m_1-m_2)^2-q^2|}f(m_1^2,m_2^2,q^2),
\end{eqnarray}
where
\begin{equation}
f(m_1^2,m_2^2,q^2)=\left\{
\begin{array}{ll}
\log\frac{\sqrt{(m_1+m_2)^2-q^2}-\sqrt{(m_1-m_2)^2-q^2}}{\sqrt{(m_1+m_2)^2-q^2}+\sqrt{(m_1-m_2)^2-q^2}}
&\quad q^2\leq(m_1-m_2)^2\\
2\arctan\sqrt{\frac{q^2-(m_1-m_2)^2}{(m_1+m_2)^2-q^2}}
&\quad (m_1-m_2)^2<q^2<(m_1+m_2)^2\\
\log\frac{\sqrt{q^2-(m_1-m_2)^2}+\sqrt{q^2-(m_1+m_2)^2}}{\sqrt{q^2-(m_1-m_2)^2}-\sqrt{q^2-(m_1+m_2)^2}}
&\quad q^2\geq(m_1+m_2)^2\\
\end{array}\right.,
\end{equation}
and
\begin{eqnarray}
b_1(m_1^2,m_2^2,q^2)&=&-\frac{1}{2}\left[\frac{m_1^2}{q^2}\left(\log\frac{m_1^2}{\mu^2}-1\right)
-\frac{m_2^2}{q^2}\left(\log\frac{m_2^2}{\mu^2}-1\right)\right]
+\frac{1}{2}\frac{m_1^2-m_2^2+q^2}{q^2}b_0(m_1,m_2,q),
\\
b_2(m_1^2,m_2^2,q^2)&=&\frac{1}{18}+\frac{1}{6}\left[
\frac{m_1^2(2m_1^2-2m_2^2-q^2)}{(q^2)^2}\log\frac{m_1^2}{\mu^2}
+\frac{m_2^2(2m_2^2-2m_1^2-q^2)}{(q^2)^2}\log\frac{m_2^2}{\mu^2}\right]
\nonumber\\&&
-\frac{1}{3}\left(\frac{m_1^2-m_2^2}{q^2}\right)^2
-\frac{1}{6}\left[
2\left(\frac{m_1^2-m_2^2}{q^2}\right)^2-\left(\frac{m_1^2+m_2^2+q^2}{q^2}\right)
\right]b_0(m_1,m_2,q).
\end{eqnarray}


\section{The matrix $M_{ij}$ and the best fit values $\hat{C}_i$}
\label{app:b}
The matrix $M_{ij}$ and the best fit values $\hat{C}_i$ in Eq.~(\ref{chi22})
are given by
\begin{eqnarray}
M&=&\frac{(1\ {\rm TeV})^{4}}{\Lambda^4}\times 10^{-2}\times
\nonumber\\
&&
\!\!\!\!\!\!\!\!\!\!\!\!\!\!\!\!\!\!\!\!
\!\!\!\!\!\!\!\!\!\!\!\!\!\!\!\!\!\!\!\!
\!\!\!\!\!\!\!\!\!\!\!
\left(
\begin{array}{c|cccccccccc}
&
C_{WB}& C_\phi^{(3)}& C_{\phi q}^{(3)}& C_{\phi q}^{(1)}&
C_{\phi t}& C_{\phi b}& C_{tW}& C_{bW}& C_{tB}& C_{bB}
\\\hline
O_{WB} & +1.10e7 & +3.06e6 & -3.16e5 & +5.47e4 & -2.70e5 & -6.16e3 & -1.35e5 & +3.11e3 & -1.71e5 & -1.40e4 \\
O_{\phi}^{(3)} & +3.06e6 & +1.07e6 & -1.40e5 & -1.03e4 & -9.49e4 & +9.46e3 & -3.39e4 & +4.04e2 & -4.77e4 & -3.85e3 \\
O_{\phi{q}}^{(3)} & -3.16e5 & -1.40e5 & +2.58e5 & +2.40e5 & +1.28e4 & -4.55e4 & +3.99e3 & -4.49e1 & +4.96e3 & +4.35e2 \\
O_{\phi{q}}^{(1)} & +5.47e4 & -1.03e4 & +2.40e5 & +2.39e5 & +1.16e3 & -4.42e4 & -1.28e2 & +3.21e0 & -8.20e2 & -3.34e1 \\
O_{\phi{t}} & -2.70e5 & -9.49e4 & +1.28e4 & +1.16e3 & +8.49e3 & -9.17e2 & +2.98e3 & -3.34e1 & +4.21e3 & +3.40e2 \\
O_{\phi{b}} & -6.16e3 & +9.46e3 & -4.55e4 & -4.42e4 & -9.17e2 & +9.83e3 & +1.13e2 & -1.46e1 & +9.24e1 & +3.20e0 \\
O_{tW} & -1.35e5 & -3.39e4 & +3.99e3 & -1.28e2 & +2.98e3 & +1.13e2 & +1.78e3 & -5.16e1 & +2.11e3 & +1.76e2 \\
O_{bW} & +3.11e3 & +4.04e2 & -4.49e1 & +3.21e0 & -3.34e1 & -1.46e1 & -5.16e1 & +2.49e0 & -4.89e1 & -4.42e0 \\
O_{tB} & -1.71e5 & -4.77e4 & +4.96e3 & -8.20e2 & +4.21e3 & +9.24e1 & +2.11e3 & -4.89e1 & +2.67e3 & +2.19e2 \\
O_{bB} & -1.40e4 & -3.85e3 & +4.35e2 & -3.34e1 & +3.40e2 & +3.20e0 & +1.76e2 & -4.42e0 & +2.19e2 & +1.82e1
\end{array}\right)
\nonumber\\
\end{eqnarray}
and
\begin{equation}
\begin{array}{c|cccccccccc}
C_i&
C_{WB}& C_\phi^{(3)}& C_{\phi q}^{(3)}& C_{\phi q}^{(1)}&
C_{\phi t}& C_{\phi b}& C_{tW}& C_{bW}& C_{tB}& C_{bB}
\\\hline
\hat{C}_i/\Lambda^2&
+0.93 & -1.63 & -689 & +689 & +939 & -0.60 & +149 & +53.9 & -78.3 & +60.3
\end{array}
\end{equation}
in units of ${\rm TeV}^{-2}$.

The numerical values of $\hat{C}_i$ depend on both the experimental values and the SM predictions.
The matrix $M$ is symmetric and positive definite, and its value
only depends on the errors of different measurements.
If any of the SM input parameters changes, the best values $\hat{C}_i$ will be affected,
but the matrix $M$ will not. The sizes of the one-sigma bounds on the operators
only depend on the matrix $M$.


\end{document}